\documentclass[journal]{IEEEtran}

\usepackage{graphicx}
\usepackage{hyperref}
\usepackage{xcolor}
\hypersetup{
  colorlinks   = true, 
  urlcolor     = red, 
  linkcolor    = blue, 
  citecolor   = blue 
}
\usepackage[misc]{ifsym}
\usepackage{multirow}
\usepackage{adjustbox}
\usepackage{xcolor}
\usepackage{subfigure}
\usepackage{amsmath}
\usepackage{subfigure}
\usepackage{pbox}
\usepackage{pifont}
\usepackage{amssymb}
\usepackage{xcolor}
\usepackage[ruled,vlined]{algorithm2e}
\newcommand{\argmin}{\arg\!\min}

\newcommand{\R}{\mathbb{R}}
\usepackage{tabu}
\usepackage[normalem]{ulem} 
\usepackage[mathlines,switch]{lineno}

\ifCLASSINFOpdf
\else
\fi
\DeclareUnicodeCharacter{2212}{-}

\begin{document}
\bstctlcite{IEEEexample:BSTcontrol}
%
\title{Spatio-temporal Multi-task Learning for Cardiac MRI Left Ventricle Quantification}

%
%
%

\author{Sulaiman Vesal,
        Mingxuan Gu,
        Andreas Maier~\IEEEmembership{Member,~IEEE},
        and Nishant Ravikumar
\thanks{S. Vesal, M. Gu, and A. Maier are with the Pattern Recognition Lab, Friedrich-Alexander-University Erlangen-Nuremberg, Germany. (E-mail: sulaiman.vesal@fau.de)}
\thanks{N. Ravikumar is with CISTIB, Centre for Computational Imaging and Simulation Technologies in Biomedicine, School of Computing, LICAMM Leeds Institute of Cardiovascular and Metabolic Medicine, School of Medicine, University of Leeds, United Kingdom.}
\thanks{The work described in this paper was partially supported by the project EFI-BIG-THERA: Integrative ‘BigData Modeling’ for the development of novel therapeutic approaches for breast cancer. The authors would also like to thank NVIDIA for donating a Titan X-Pascal GPU.

Copyright $\copyright$ 2020  IEEE.  Personal  use  of  this  material  is  permitted. However,  permission  to  use  this  material  for  any  other  purposes  must  be obtained from the IEEE by sending a request to pubs-permissions@ieee.org.
}
}
%
%

\markboth{Accepted at IEEE Journal of Biomedical and Health Informatics. \MakeLowercase {Final version available at: 
\href{https://doi.org/10.1109/JBHI.2020.3046449}{https://doi.org/10.1109/JBHI.2020.3046449}}}%
{Shell \MakeLowercase{\textit{\textit{et al.}}}: Bare Demo of IEEEtran.cls for IEEE Journals}
%



\maketitle
\begin{abstract}
Quantitative assessment of cardiac left ventricle (LV) morphology is essential to assess cardiac function and improve the diagnosis of different cardiovascular diseases. In current clinical practice, LV quantification depends on the measurement of myocardial shape indices, which is usually achieved by manual contouring of the endo- and epicardial. However, this process subjected to inter and intra-observer variability, and it is a time-consuming and tedious task. In this paper, we propose a spatio-temporal multi-task learning approach to obtain a complete set of measurements quantifying cardiac LV morphology, regional-wall thickness (RWT), and additionally detecting the cardiac phase cycle (systole and diastole) for a given 3D Cine-magnetic resonance (MR) image sequence. We first segment cardiac LVs using an encoder-decoder network and then introduce a multitask framework to regress 11 LV indices and classify the cardiac phase, as parallel tasks during model optimization. The proposed deep learning model is based on the 3D spatio-temporal convolutions, which extract spatial and temporal features from MR images.
We demonstrate the efficacy of the proposed method using cine-MR sequences of 145 subjects and comparing the performance with other state-of-the-art quantification methods. The proposed method obtained high prediction accuracy, with an average mean absolute error (MAE) of 129 $mm^2$, 1.23 $mm$, 1.76 $mm$, Pearson correlation coefficient (PCC) of 96.4\%, 87.2\%, and 97.5\%  for LV and myocardium (Myo) cavity regions, 6 RWTs, 3 LV dimensions, and an error rate of 9.0\% for phase classification. The experimental results highlight the robustness of the proposed method, despite varying degrees of cardiac morphology, image appearance, and low contrast in the cardiac MR sequences. 


\end{abstract}

\begin{IEEEkeywords}
Left Ventricle Quantification, Cardiac MRI, Cardiac Segmentation, Deep Learning, Myocardial Infraction
\end{IEEEkeywords}

%
\IEEEpeerreviewmaketitle

\section{Introduction}
Cardiovascular diseases (CVDs) and other cardiac pathologies are the leading cause of death worldwide \cite{10.1093/eurheartj/ehx628}, \cite{doi:10.1161/CIR.0000000000000485}, \cite{doi:10.1177/2048004016687211}. Timely diagnosis is crucial for improving survival rates and delivering high-quality patient care. Cardiac magnetic resonance imaging (MRI) is a non-invasive imaging modality used to detect and monitor cardiovascular diseases. Quantitative assessment and analysis of cardiac-MR images are indispensable for diagnosis and devising suitable treatments. The reliability of quantitative metrics that characterize cardiac function such as myocardial deformation and ventricular ejection fraction are heavily dependent on the precision of ventricle quantification \cite{doi:10.1002/mrm.26631}. 

In everyday clinical practice, evaluation of LV function is often conducted by visual assessment and semi-automatic tools to quantify dynamics in MRI \cite{KIM20091}, \cite{doi:10.1161/CIRCIMAGING.117.007165} \cite{Bai2018}. Hence, clinical evaluation of regional LV function is mostly qualitative and by visually observing myocardial wall displacement and deformation. Naturally, this process can be error-prone either due to artifacts arising from cardiac, respiratory or patient motion, variations in image contrast, or human error. This may prevent an accurate evaluation of LV structures. On the other hand, LV assessment by cardiologists requires extensive expertise and experience \cite{kurzendorfer}, \cite{10.1007/978-3-642-33418-4_66}, \cite{10.1093/ehjci/jev014}. A central part of morphological cardiac quantification involves manual/semi-automatic contouring of the endo- and epicardial walls of the left ventricular myocardium. It is time-consuming and often subjected to high intra and inter-observer variability. Moreover, the myocardium contouring process is performed on the end-systolic (ES), and end-diastolic (ED) frames that are inadequate for comprehensive analysis of heart function (across the full cardiac phase) \cite{KIM20091}. Notwithstanding recent advances, LV segmentation is still a challenging problem due to limited contrast between tissue boundaries, and pathology-driven variability in shape and appearance in cine-MR sequences. \cite{6650070}, \cite{Tao2018}.

\begin{figure}[!t]
    \centering
    \includegraphics[width=0.48\textwidth]{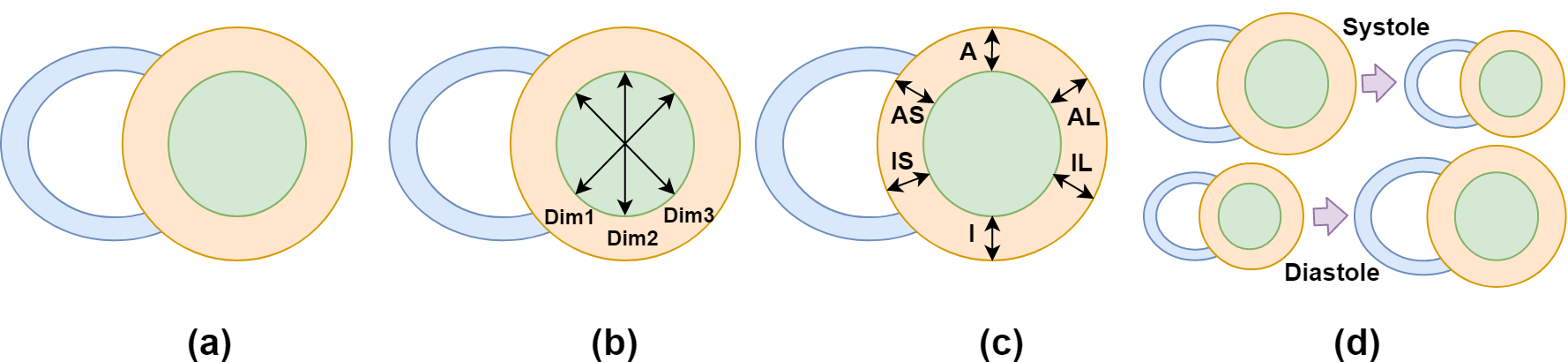}
    \caption{A schematic representation of LV indices for short-axis cardiac Cine-MR image. The LV and Myo cavity areas are shown with green and blue colors in (a) and three LV cavity dimensions with black arrows in (b). (c) shows Six myocardial regional-wall thicknesses, namely anterolateral (AL), inferolateral (IL), inferior (I), inferoseptal (IS),  anterior (A), inferoseptal (IS) and anteroseptal (AS). The cardiac phase (systole or diastole) is shown in (d).}
    \label{fig:indices}
\end{figure}

To address these challenges, we propose a deep learning approach to enable automatic full quantification of LV morphology in short-axis cardiac cine-MR images without any further information. We investigate the use of temporal and spatial information to estimate the cardiac phase, diameters of the LV blood pool (or cavity) along with different directions, regional wall thicknesses (RWTs) (as depicted in Fig. \ref{fig:indices}), and LV cavity and myocardial areas. Comprehensive assessment of these measures requires analysis of images from the entire $2D+t$ cine-MR sequence (covering the full cardiac cycle), thereby considering the temporal dynamics of the heart. \\

\IEEEpubidadjcol

\section{Related Work}

In recent years, many studies have focused on automatic full cardiac LV morphological quantification. These methods are designed either as multi-stage or end-to-end approaches, in terms of their training strategy. Traditional multi-stage methods are mainly based on myocardium segmentation \cite{8674807}, \cite{Suinesiaputra2015}, \cite{RUIJSINK2019}, \cite{ATTAR201926}, where first the LV endocardium and epicardium are segmented and then the desired LV indices are estimated based on segmentation masks. The latter takes advantage of machine learning algorithms \cite{BENAYED201287}, \cite{ZHEN2016120}, where features are extracted automatically from cardiac MR images, and a regression model utilizes these features to estimate the LV indices. In comparison to the multi-stage methods, end-to-end methods \cite{8674807}, \cite{indnet}, \cite{10.1007/978-3-319-59050-9_40} combine feature extraction and regression together using deep neural networks. 

One of the earliest works for LV quantification based on manual segmentation proposed by Suinesiaputra \textit{et al.} \cite{Suinesiaputra2015}. They asked seven cardiologists to manually delineate contours around myocardium and LV cavity volume at the ES and ED phases to evaluate cardiac function. As we know, manual segmentation is very time-consuming, subjective, and not very efficient. To tackle these limitations, several automatic segmentation algorithms \cite{661186}, \cite{spatio}, \cite{10.1007/978-3-030-12029-0_41}, \cite{XUE201854} have been proposed. Wang \textit{et al.} \cite{6708423} considered a Bayesian method for two-ventricular volume estimation that used a likelihood function for exploiting appearance features and a probability model to incorporate the area correlation between the cavities. Zhen \textit{et al.} \cite{ZHEN2016120} initially extracted hierarchical profiles using multi-scale deep neural networks and then placed them in a random regression forest to estimate the left ventricle. In this type of two-step approach, there is only a forward linkage and no feedback from the second step. Therefore the features extracted in the first stage may not be closely related to the target tasks, and the estimated results in the second stage may not be very accurate.

Moreover, the direct regression-based methods have been also used to quantify LV indices either in an end-to-end or multi-stage fashion. Xue \textit{et al.} \cite{XUE20182} proposed an integrated model to present multiple LV criteria, including two regions and six RWTs per frame within the cardiac cycle. To model the temporal dynamics of cardiac sequences, they employed a Recurrent Neural Network (RNN) followed by a Convolutional Neural Network (CNN) module to regress six RWTs \cite{indnet}.Additionally, Xue \textit{et al.} \cite{10.1007/978-3-319-59050-9_40} focused on the quantification of complete LV measurements, which requires estimating the regions, orientation dimensions, and RWTs simultaneously for each MR image. To improve the prediction accuracy, they used both CNN and RNN modules and modeled the correlations between the different LV criteria with a multi-task loss. Nevertheless, these methods don't process the whole cardiac cycle as a whole for feature extraction, but rather an embedding of 3-5 neighboring MR frames to incorporate temporal information. Therefore, they do not guarantee the temporal dynamic consistency of the estimated volumes across the whole cardiac cycle. Wang \textit{et al.} \cite{8674807} proposed a cascaded segmentation and regression network, in which the segmentation component extracts left ventricular myocardial contours, and the regression component estimates the desired LV criteria. However, this method only computes the LV indices and not the cardiac phase cycle. Another study \cite{spatio} processes stacks of adjacent slices($k=5$) using 3D convolutional kernels to incorporate temporal information within the learned model. Tao \textit{et al.} \cite{Tao2018} in a more clinically adapted approach tested three different CNNs for fully automated quantification of LV on multi-centers and multi-vendors study. However, this work was performed on retrospective data and not cover a wide range of cardiovascular abnormalities, which is clinically more demanding. Recently, disentanglement representation learning methods also investigated \cite{Meng2019} \cite{CHARTSIAS2019101535} to extract generalize features within a multi-task framework. The model encodes informative features for different tasks and employing the adversarial regularization to enforce the extracted features to be minimally informative about irrelevant tasks.

Inspired by previous works, and to address the challenges as yet unmet by current methods for LV quantification, we propose a novel end-to-end multi-task learning framework based on $2D + t$ spatio-temporal convolutions to simultaneously tackle multiple tasks and allow them mutually learn from each other \cite{ZHANG201910}, \cite{NIPS2004_2638},\cite{Maier2019}. The proposed approach permits accurate quantification of standard LV indices and provides 3D segmentation for the blood pool and myocardium of the LV for further morphological analysis. Introducing a single model that is capable of solving multiple tasks at the same time can be clinically very relevant and reduce the overhead for the cardiologists to have individual models for each sub-task including ventricle segmentation and quantification.
 \\

\noindent In summary, our main contributions are three folds: 
\begin{itemize}
    \item First, we propose an end-to-end deep learning model that directly learns temporal and spatial features using 3D spatio-temporal convolutions from the estimated 3D cine-MR segmentation masks. The proposed model takes the full temporal cine-MR sequence into account, to quantify the LV, rather than a single 2D image or by concatenating a few 2D image slices from adjacent time frames (i.e. 2.5D).
    \item Second, a multi-task network is introduced to leverage the shared information useful for LV segmentation, LV indices regression, and cardiac phase cycle classification tasks, by jointly optimise all three. We further demonstrate with empirical evidence that the temporal information and volumetric quantification improves prediction accuracy significantly compared to 2D and 2.5D deep learning models.
    \item Third, we validate our proposed method using the publicly available LVQuan 2018 benchmark dataset, which provides short-axis cine-MR sequences with annotations for the above indices (for the whole cardiac cycle). The proposed method achieved better robustness and interpretation for LV quantification and morphology assessment in comparison to state-of-the-art methods. The code and model are available at: \url{https://github.com/sulaimanvesal/CardiacQuanNet}
\end{itemize}



\section{Materials and Methods}
\begin{figure*}[!t]
    \centering
    \includegraphics[width=15.8cm]{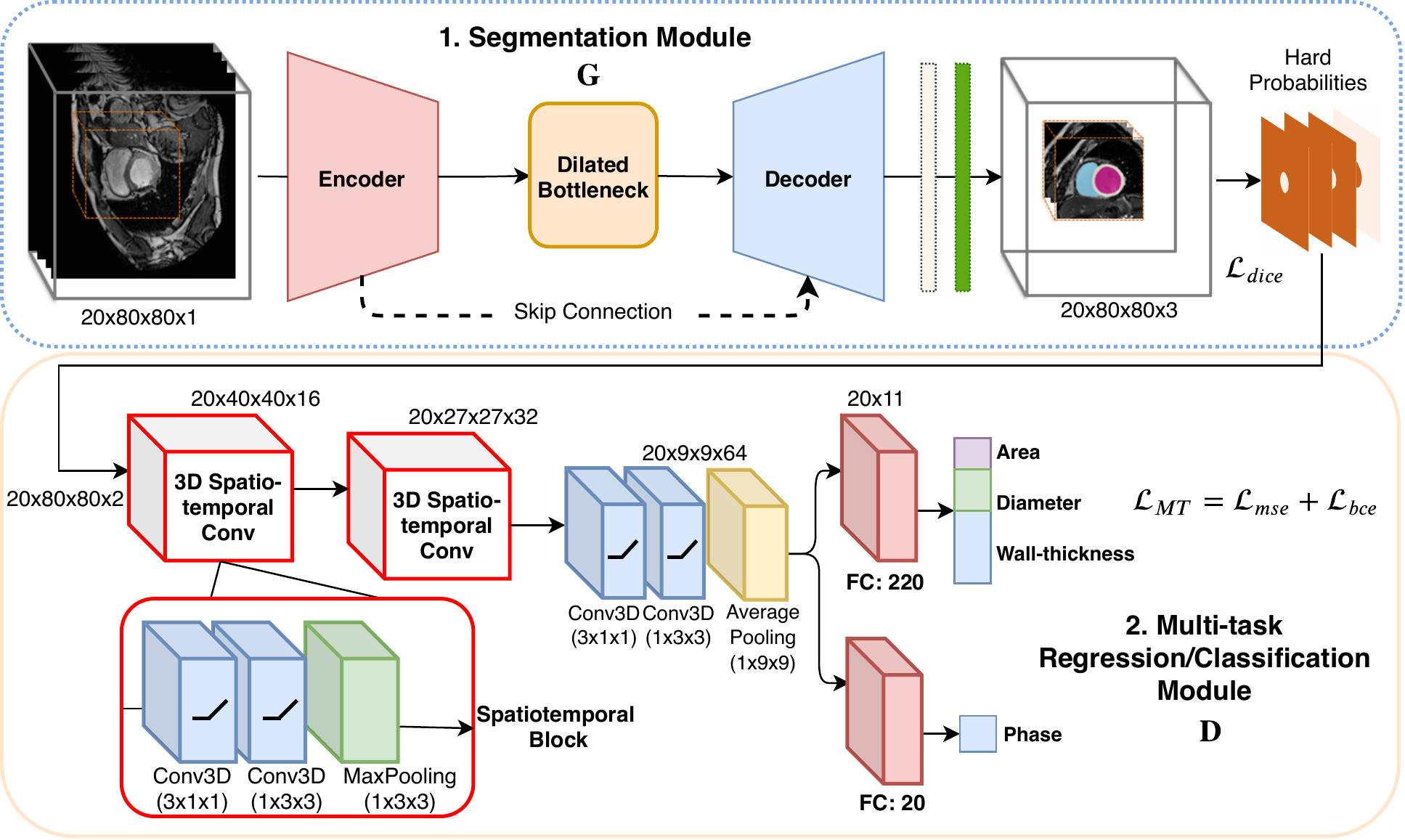}
    \caption{Network architecture overview. Given Cine-MR volumes with the size $t \times h \times w$ from the training dataset as input, we pass it through the segmentation network to obtain output segmentation masks for the LV cavity and Myo. A segmentation loss is computed based on the ground truth. To make predictions for 11 indices and cardiac phase detection, we utilize a multi-task spatio-temporal network with two parallel branches. Then a multi-task loss is calculated on the target prediction for both regression and classification tasks and is back-propagated to the segmentation network.}
    \label{fig:worflow}
\end{figure*}
In this section, we first describe the details about the dataset. Further, we introduce our proposed Spaito-temporal multi-task learning network pipeline and its component for LV segmentation, regression, and classification. Eventually, the objective function and training settings are described.  

\subsection{Dataset}
We validated our proposed framework on the STACOM LVQuan 2018 challenge training dataset \cite{XUE20182}. The dataset collected from 3 different hospitals and in collaboration with two healthcare centers, namely London Healthcare Center and St. Josephs Healthcare \cite{indnet},  \cite{XUE201854}. It consists of 2D Cine-MR images of $n_{S} = 145$ patients with an average age of 58.9 years. The Cine-MR images have a pixel spacing ranges between 2.0833 mm/pixel and 0.6836 mm/pixel. The dataset has a set of various pathologies like myocardial hypertrophy, regional-wall motion abnormalities, atrial septal defect, mildly enlarged LV, LV dysfunction, etc. Each Cine-MR sequence has 20 frames per cycle, resulting in a total of 2900 images in the training dataset. Following the standard American Heart Association (AHA) recommendation \cite{doi:10.1161/hc0402.102975}, each frame in the dataset includes only mid-cavity regions, which is perpendicular to the long axis of the heart.

All cardiac images annotated manually to obtain the epicardium and endocardium boundaries, which are double-checked by two experienced cardiologists. The ground truth values of LV indices are computed based on these delineations. The RWTs and LV blood pool dimensions indices are normalized with respect to the image size, while the areas normalized by the number of pixels (2900). After the training step, the computed indices are converted back to physical thickness ($mm$) and area ($mm^2$) by changing the resizing procedure and multiplying each subject with their corresponding pixel spacing. To evaluate and compare model performance, we employed five-fold cross-validation similar to other studies. Off-line data augmentation was used by randomly rotating, flipping horizontally/vertically, and applying elastic deformation to the training images. This process increased the number of training samples in each fold by a factor of 8.

\subsection{Network Architecture}
As discussed previously, our LV quantification framework consists of two modules: a segmentation network $\mathbf{G}$ and the multi-task classification/regression network $\mathbf{D}$. During training, we first provide an MR image sequence $\mathcal{I}_{S} \in \R^{t \times h \times w \times 1}$ (with annotations) to the segmentation network for optimizing $\mathbf{G}$. Then, we convert the soft probabilities of the \textit{softmax} layer to hard probabilities, and provide the LV and Myo segmentation predictions ($\mathcal{P}_{S} \in \R^{t \times h \times w \times 2}$) as the input to $\mathbf{D}$. 
Here, $2$ refers to image channels corresponding to the LV and Myo segmentation masks, and we discard the background channel as the indices are computed from LV and Myo channels only. The network propagates gradients from $\mathbf{D}$ to $\mathbf{G} $, which in turn encourages $\mathbf{G}$ to optimize its weights with respect to both the segmentation labels (tissue boundaries) and the LV indices of interest. Fig. \ref{fig:worflow} shows an overview of the proposed algorithm. In this section, we first describe the left ventricle segmentation module and subsequently, the multi-task network for regression and classification of 11 indices and cardiac phase recognition, respectively. The $\mathbf{G}$ and $\mathbf{D}$ modules are trained using two strategies: (1) multi-stage and (2) end-to-end. In an end-to-end fashion, we optimize both networks simultaneously.  

\subsection{Left Ventricle Segmentation}
\begin{figure}[!t]
    \centering
    \includegraphics[width=\columnwidth]{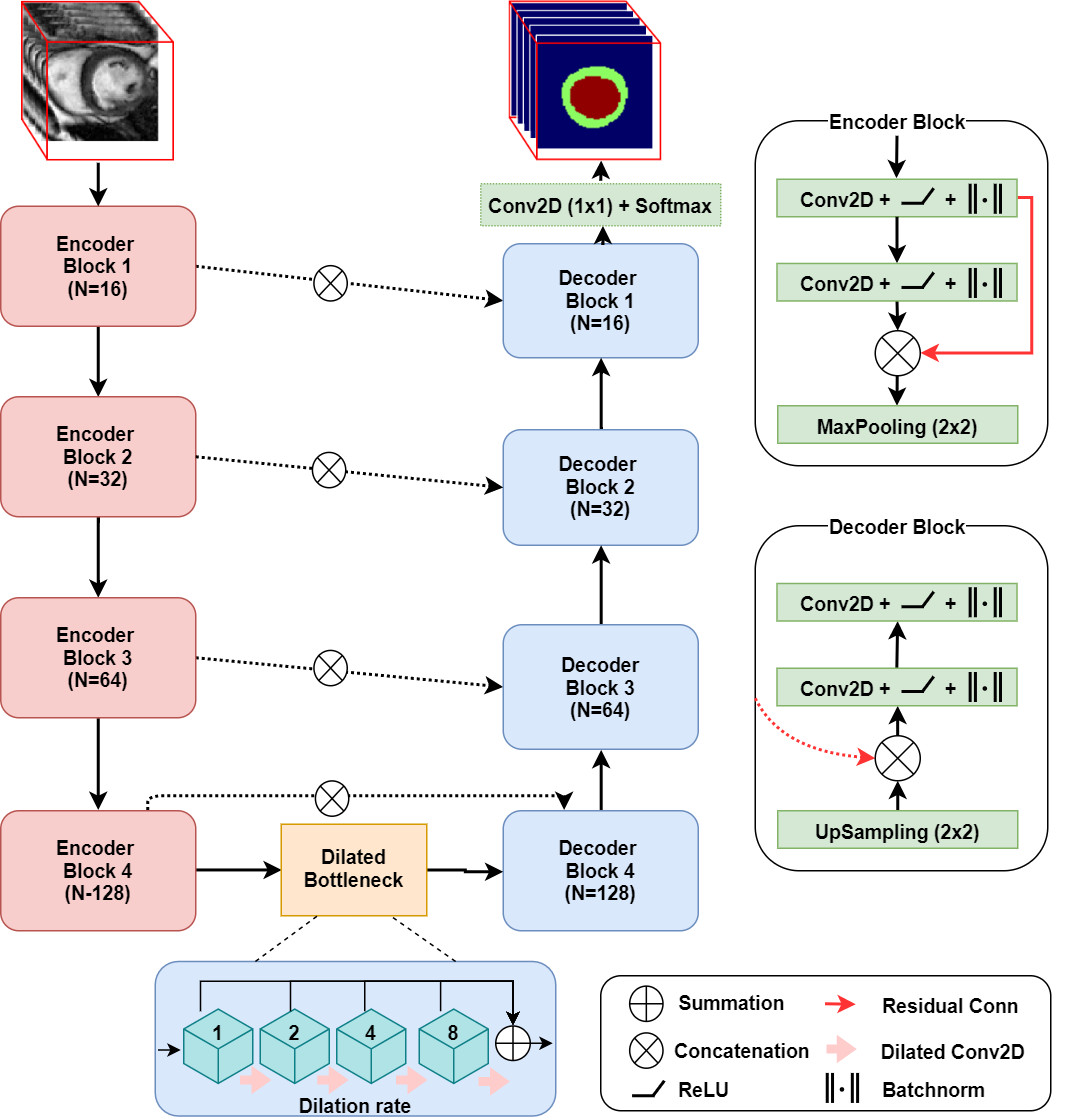}
    \caption{The schematic illustration of our left ventricle segmentation network equipped with dilated convolutions(4 dilation rate 1-8) in the bottleneck to capture multi-scale features and residual blocks in the encoder path.}
    \label{fig:unet}
\end{figure}

To segment the LV blood pool and myocardium of the LV, we employ a fully convolutional network architecture inspired by \cite{10.1007/978-3-030-12029-0_35} called Dilated Residual-UNet (DR-UNet), which is depicted in Fig. \ref{fig:unet}. The segmentation network $\mathbf{G}$ has an encoder and decoder paths that are connected by a bottleneck block.  Every block in encoder and decoder paths has two 2D convolution layers followed by a Rectified Linear Unit (ReLU), batch-normalization, and a 2D max-pooling layer to reduce the dimensions of feature maps. To improve the flow of gradients, and enforcing the encoder to extract more discriminative features, a residual connection \cite{resDeep} added in each encoder block. The last layer of the network has a \textit{softmax} activation function to produce probability segmentation maps for each class. In DR-UNet, the normal convolution layers are replaced with dilated convolutions to permit the network to capture both global and local contextual information by increasing the respective field. A sequence of dilated convolutions can introduce gridding effects (different output nodes use disjoint subsets of input nodes) if dilation rates are not selected properly \cite{wang2018smoothed}. As a consequence, we have created a block of stacked dilated convolutions whose outputs are summed together. This way, each subsequent layer has full access to previous features learned using different dilation rates, addressing the issue of gridding artefacts. In our network settings, we used four dilated convolutions with a dilation rate of $1-8$ in the network bottleneck.
In the case of the end-to-end training strategy, we adapted the segmentation network $\mathbf{G}$ from 2D to 3D by replacing all the 2D convolution operations with 3D convolution layers, while the rest of the network remained the same. It is because we consider the full temporal sequence as the input for the regression and classification task, resulting in an input tensor size of 20$\times$80$\times$80$\times$2. On the other hand, the 3D spatio-temporal module for classification and regression have 3D kernels which required 3D input. Therefore, we selected DR-UNet 3D as the segmentation backbone.

\subsection{Left Ventricle Quantification}
To quantify the cardiac LV, we propose a multi-task CNN architecture, which is trained both in an end-to-end and multi-stage fashion. This network employs spatio-temporal convolutions to consider not only spatial but also temporal information. 3D CNNs applied to $2D + t$ image frames to preserve temporal information and propagate it through the layers of the network \cite{8578773}, \cite{10.1007/978-3-642-15567-3_11}. The cardiac Cine-MR dataset has a short-term temporal dynamic between neighboring slices in the sequence over one cardiac cycle. For this reason, we consider 3D information as features for temporal modeling on the $2D + t$ MR images. The 3D convolution addresses each image with assistance from adjacent slices, and it can represent more robust structural features as well as temporal information \cite{univis91902143}.

Fig. \ref{fig:worflow} illustrates our proposed spatio-temporal network architecture. The proposed network consists of three spatio-temporal blocks and two task-specific parallel branches. The first branch computes 11 indices for LV areas, LV dimensions, and RWTs, and the second branch classifies the cardiac phase across entire sequences. Cardiac phase-detection is normally considered as a sequence modeling task as the temporal dynamics are important for determining the cardiac phase. RNN blocks are widely used for this type of task \cite{XUE201854}, but these models are difficult to optimize. The proposed spatio-temporal blocks on the other hand, already take into account both the spatial and temporal dynamics of slices, which removes the need for RNN blocks.

The input to the encoder part has a size of $t \times h \times w$, where $t$ is the temporal axis, and $h \times w$ are the spatial axes. Each spatio-temporal block has two 3D convolution layers and a subsequent 3D MaxPooling layer. The first 3D convolution layer has a kernel size of 3$\times$1$\times$1 that captures temporal features across the time axis. The second convolution layer extracts spatial features with a kernel size of 1$\times$3$\times$3 and strides of 1. The MaxPooling layer has a window size of 1$\times$3$\times$3 as we only want to downsample the inputs along the spatial dimensions while keeping the number of frames fixed (as we would like to compute LV indices for each frame). In comparison to fully 3D convolution operation, this decomposition offers two advantages as highlighted in \cite{8578773}. First, in this setup, the number of parameters is not changed. However, it increases the number of nonlinearities in the network due to the additional ReLU between the two convolution layers in each block. Doubling the number of nonlinearities enhances the complexity of functions, which approximate the effect of a big filter by applying multiple smaller filters with additional nonlinearities in between. The second benefit is that forcing the 3D convolution into separate spatial and temporal components makes the optimization easier \cite{8578773}. This in turn helps reduce the error rate compared to conventional 3D CNNs of the same capacity \cite{8578773}. For all convolution layers, we initialize the kernels with the He initializer \cite{He} and employ $\mathcal{L}_{2}$ weight regularization to reduce the overfitting of the proposed model on the training data.

\subsection{Pre-processing} \label{secpre} 
\begin{figure}[!t]
    \centering
    \includegraphics[width=\columnwidth]{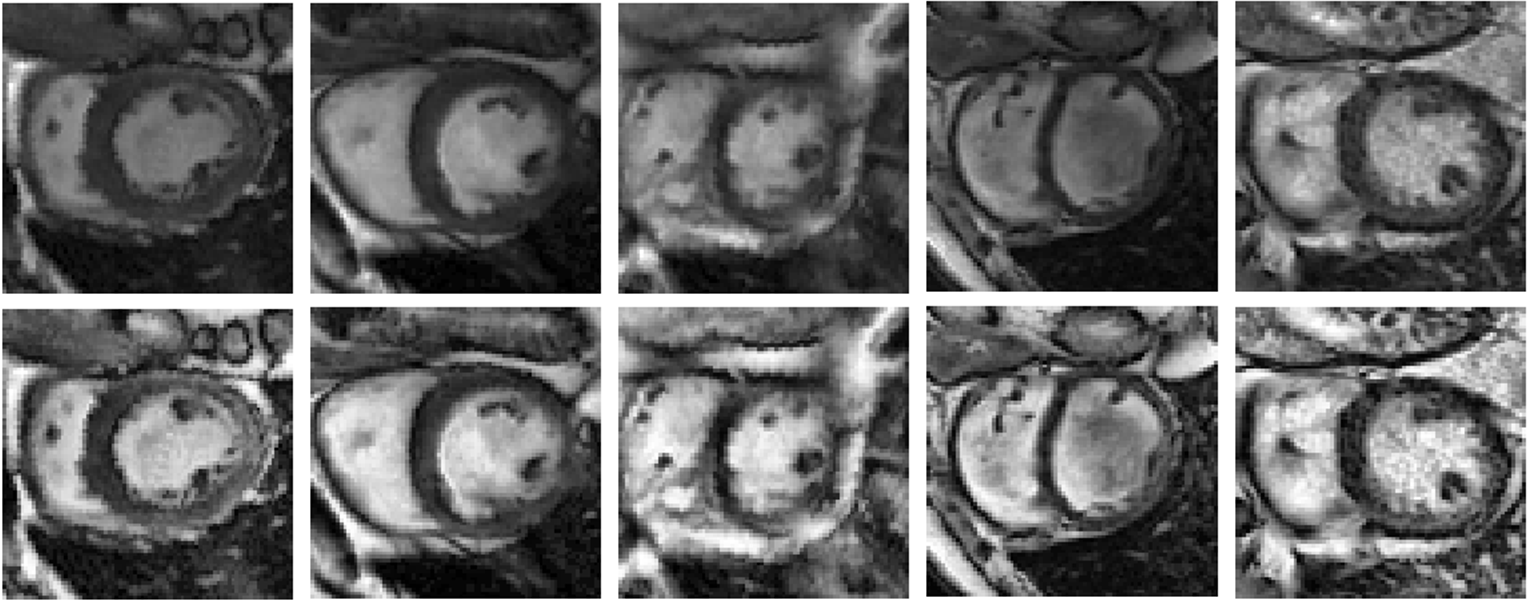}
    \caption{LV MR image slices before(first row) and after pre-processing and normalisation (second row).}
    \label{fig:preproc}
\end{figure}

All cardiac images were preprocessed by the challenge organizer, including landmark labeling to find the ROI, rotation to align the volumes, ROI cropping, and resizing. The resulting images are 80 $\times$ 80 in size. The LVQuan 2018 dataset images vary a lot in terms of contrast and brightness. The variability results from different acquisition parameters and scanners are always a challenge for designing a robust and generalized neural network model. Contrast limited adaptive histogram equalization (CLAHE) is applied to further improve the contrast of Cine-MR images as well as reducing the variability across the dataset, particularly for those images with low contrast \cite{Zuiderveld}. Subsequently, the images were normalized by subtracting the mean and dividing by the standard deviation for each sequence. These prepossessing steps significantly improved the segmentation accuracy for DR-UNet. Fig. \ref{fig:preproc} presents a few sample images before and after employing CLAHE and image normalization. From these images, it is evident that variability in brightness and contrast is largely diminished following contrast enhancement.

\noindent\textbf{Target Labels Normalisation:} In the LVQuan 2018 dataset, there is a large amount of variation in the magnitude of the various indices. Different distributions of different target indices can cause unbalanced and unstable training. To tackle this issue, we normalize the label indices using the $z$-score for all the 11 indices (2 LV and Myo areas, 3 LV blood pool dimensions, and 6 RWTs) across the whole dataset. After training for the final evaluation, we scale the indices back to the original value by multiplying the standard deviation and adding the mean value. It should be noted that this normalization step is only for quantification labels, while the image normalization process described in the preprocessing section. 

\subsection{Loss Functions}
\noindent \textbf{Segmentation Loss:} The segmentation network $\mathbf{G}$ is trained with a multi-class soft-Dice loss, which shown to be less sensitive when there is a huge class imbalance within the dataset in comparison to binary cross-entropy loss.  Many recent studies \cite{10.1007/978-3-030-12029-0_35}, \cite{7785132} also used this objective function for medical image segmentation. We first compute the Dice loss for every class individually and then average it over the number of available classes. To segment a Cine-MR image $\mathcal{I}_{S} \in \R^{t \times h \times w \times 1}$ with having LV, Myo and background as labels, the output of \textit{Softmax} layer is three probability maps for classes $k = 0, 1, 2$ where for each pixel $\sum_{c}\pmb{y}_{n,k} = 1$. Given the ground-truth label $\pmb{\hat{y}}_{n,k}$ for that identical pixel, the multi-class soft Dice loss is computed as follows:
\begin{equation}
\label{eq1lo}
	\mathcal{L}_{dice}(\pmb{y}, \pmb{\hat{y}})  = 1- \frac{1}{K}(\sum_{k}\pmb{w}_{k}\frac{\sum_{n}\pmb{y}_{nk} \pmb{\hat{y}}_{nk}}{\sum_{n}\pmb{y}_{nk} + \sum_{n}\pmb{\hat{y}}_{nk}})
\end{equation}
where $\pmb{w}_{k}$ is the weight factor to tackle class imbalance as Myo region has fewer pixels compared to the other two classes. We empirically set the weights for each class: $\{BG: 0.2, LV: 0.3, Myo: 0.5\}$. We achieved better segmentation performance by weighting the Myo class higher.
\\

\noindent \textbf{Classification Loss:} Given the class probability output $\mathcal{P}_{phase}=\mathbf{D}(\mathcal{I}_{S})$ from the cardiac phase classification branch, the cross-entropy loss $\mathcal{L}_{bce}(\pmb{y}, \pmb{\hat{y}})$ for the two classes (i.e., ED and ES) can be written as:
\begin{equation}
\begin{split}
\mathcal{L}_{bce}(\pmb{y}, \pmb{\hat{y}}) = -\frac{1}{N}\sum_{i=1}^{N}(\pmb{y}_{i}).log(\pmb{\hat{y}}_{i})  + (1-\pmb{y}_{i}).log(1-\pmb{\hat{y}}_{i})
\end{split}
\end{equation}
where, $\pmb{y}$ is the label (1 for ED phase and 0 for ES phase) and $\pmb{\hat{y}}$ is the predicted probability. 
\\

\noindent \textbf{Multi-Task Loss:} To train the multi-task regression and classification module $\mathbf{D}$, we optimized the model using a joint loss, combining $\mathcal{L}_{bce}(\pmb{y}, \pmb{\hat{y}})$ and Mean Squared Error (MSE), for cardiac phase classification and LV indices regression, respectively. The loss is formulated as:
\begin{equation}
\label{eq16}
\mathcal{L}_{mse}(\pmb{y}, \pmb{\hat{y}}) = -\frac{1}{N}\sum_{s=1}^{11}\sum_{i=1}^{N}||\pmb{y}_{s,i}-\pmb{\hat{y}}_{s,i}||_{2}^{2}
\end{equation}
\begin{equation}
\label{eq15}
	\mathcal{L}_{mt}(\pmb{y}, \pmb{\hat{y}})  = \argmin_{\hat{y}}  \lambda_{1}.\mathcal{L}_{mse}(\pmb{y}, \pmb{\hat{y}}) + \lambda_{2}.\mathcal{L}_{bce}(\pmb{y}, \pmb{\hat{y}}) 
\end{equation}
where, $\lambda_{1}$ and $\lambda_{2}$ are the weights to control the influence of the individual tasks on the combined loss. We empirically set $\lambda_{1} =1$ and $\lambda_{2}=4$. This is due to the fast convergence of $\mathcal{L}_{mse}$, which necessitated higher weights for the classification task in order to stabilize the training process. Equation (\ref{eq15}) can be extended further to train the entire pipeline including the segmentation network $\mathbf{G}$, in an end-to-end fashion:
\begin{equation}
\label{eq17}
\begin{split}
	\mathcal{L}_{mt}(\pmb{y}, \pmb{\hat{y}})  = \argmin_{\hat{y}}   \lambda_{1}.\mathcal{L}_{dice}(\pmb{y}, \pmb{\hat{y}})  + \lambda_{2}.\mathcal{L}_{mse}(\pmb{y}, \pmb{\hat{y}}) \\ + \lambda_{3}.\mathcal{L}_{bce}(\pmb{y}, \pmb{\hat{y}}) 
\end{split}
\end{equation}

\begin{figure*}[ht]
\centering
\includegraphics[width=\textwidth]{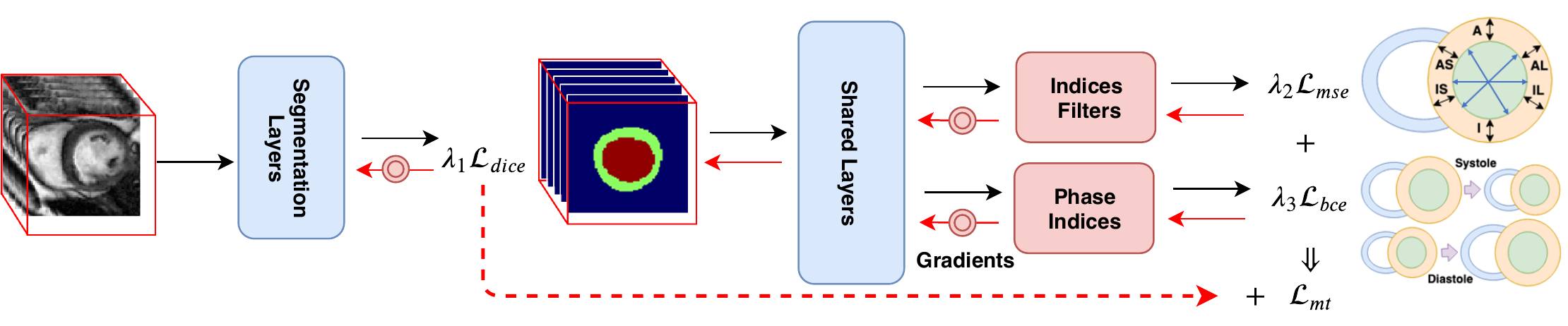}
\centering\caption{Representation of gradient flow between segmentation, regression and classification tasks in our cardiac quantification framework.}
\label{fig:app2}
\end{figure*}

Here, we combine the segmentation loss $\mathcal{L}_{dice}$ with the multi-task loss for regression and classification tasks. However, to have control over the influence of different losses and gradients in $\mathcal{L}_{mt}$, we have empirically set $\lambda_{1} =10$ , $\lambda_{2}=1$ and $\lambda_{3}=1$ as weights. Since the regression and classification part depends on the output of the segmentation network, we have given more weights to this task. This enforces the method to produce more accurate myocardial segmentation.
Fig. \ref{fig:app2} illustrates a simplified view of our framework, which shows the flow of gradients between different modules and how we train all the three tasks end-to-end.    

\subsection{Network Training}
As discussed previously, we trained our LV quantification pipeline using both multi-stage and end-to-end strategies. In each training batch, the MR sequence $\mathcal{I}_{S}$ is provided to the segmentation network $\mathbf{G}$ which is trained by optimizing $\mathcal{L}_{dice}(\pmb{y}, \pmb{\hat{y}})$ in (Eq. \ref{eq1lo}), to generate the output probability map $P_{s}$. These soft probability masks are subsequently converted to hard probabilities and passed on to the multi-task network $\mathbf{D}$. The latter in turn is trained by optimizing $\mathcal{L}_{mt}$ in (Eq \ref{eq15}). The predicted normalized LV indices are subsequently converted back to physical values before evaluation.

The performance of our framework evaluated across 5-fold cross-validation experiments. In each fold, there are 29 subjects. Four folds used to train the spatio-temporal multi-task learning model and the fifth to test. We repeated the same procedure five times until the LV indices of all subjects were obtained. The proposed deep learning model designed and developed in Keras and TensorFlow \cite{Tensoflow}, which is an open-source deep-learning library for Python. All the networks trained on an NVIDIA Titan X-Pascal GPU with 12GB memory. In the multi-stage strategy, we trained the segmentation network $\mathbf{G}$ using the ADAM optimizer \cite{Adam}. A fixed learning rate of $0.0001$ with exponential decay rates of the $1$\textsuperscript{st} is used, and Adam momentum parameters were set to $0.9$ and $0.999$, respectively. The multi-task network $\mathbf{D}$ was also trained using Adam optimizer with a learning rate of $0.004$, and with similar decay rate and momentum as $\mathbf{G}$. For the end-to-end training strategy, we employed the same optimizer and hyperparameters to make both the models comparable.

\section{Experiments and Results}

\subsection{Evaluation Metrics}
To evaluate the performance of the methods quantitatively, we used the Pearson correlation coefficient (PCC) and Mean Absolute Error (MAE) metrics. For the cardiac LV indices MAE and PCC are computed as follows:

\begin{equation}
    MAE_{ind} = \frac{1}{N}\sum_{i=1}^{N}|\pmb{\hat{y}}_{ind}^{i}-\pmb{y}_{ind}^{i}|
\end{equation}
\begin{equation}
    PCC_{ind} = \frac{\sum_{i=1}^{N}(\pmb{\hat{y}}_{ind}^{i}-\pmb{\bar{y}}_{ind}^{i})(\pmb{y}_{ind}^{i}-\pmb{\bar{y}}_{ind}^{i})}
    {\sqrt{\sum_{i=1}^{N}(\pmb{\hat{y}}_{ind}^{i}-\pmb{\hat{\bar{y}}}_{ind})^{2}(\pmb{y}_{ind}^{i}-\pmb{\bar{y}}_{ind})^{2}}}
\end{equation}

where, $ind \in (A1, A2, D1...D3, RWT1...RWT6)$, $\pmb{\hat{y}}_{ind}$ is the estimated value by the model and $\pmb{y}_{ind}$ is the ground-truth value provided by the rater. Here, $\pmb{\bar{y}}_{ind}$ and $\pmb{\hat{\bar{y}}}_{ind}$ are their mean values, respectively.

To evaluate and assess the model performance for cardiac phase classification, we used the Error Rate (ER), which is defined as:
\begin{equation}
    ER_{phase} = \frac{1}{N}\sum_{i=1}^{N}(\pmb{\hat{y}}_{phase}^{i} \neq \pmb{y}_{phase}^{i})100\%
\end{equation}
where $\pmb{y}_{phase}$ and $\pmb{\hat{y}}_{phase}$ are the ground-truth and estimated classes for the cardiac phase, respectively. To evaluate the accuracy of the segmentation results, we used the well-known metrics in the medical image segmentation field, the Dice coefficient (Dice) score, and Hausdorff distance (HD).  \cite{10.1007/978-3-030-12029-0_35}.
 \begin{table*}[ht]
   \centering
   \caption{Segmentation accuracy using different evaluation metrics and training strategies. As an ablation study, the number of filters, post-processing and loss function have been changed to evaluate the performance.Here, Connected component analysis is abbreviated as CCA.}
     \begin{tabular}{lcccccccccc}
     \hline
     \multirow{2}[4]{*}{Experiments} & \multicolumn{1}{c}{\multirow{2}[4]{*}{Filters}} & \multirow{2}[4]{*}{CCA} & \multirow{2}[4]{*}{Weighted-loss} & \multicolumn{3}{c}{Dice Score $\uparrow$} & \multicolumn{3}{c}{HD [mm] $\downarrow$} & \multirow{2}[4]{*}{Parameters}\\
\cline{5-10}        &    &    &    & Bg & LV & Myo & Bg & LV & Myo & \\
     \hline
      CSRNet \cite{8674807} & - &  -  &  - & 0.989 & 0.959 & 0.886 & 4.88 & 3.55 & 5.43 & 0.3 M \\
     \hline
      & 8  &  $\times$  &  $\times$  & 0.978 & 0.945 & 0.854 & 4.54 & 4.65 & 11.92 & 0.9 M \\
      & 16 &  $\times$  &  $\times$  & 0.982 & 0.950 & 0.868 & 4.03 & 5.76 & 11.53 & 3.6 M \\
     DR-UNet 2D & 8  &  $\checkmark$ &  $\times$  & 0.981 & 0.951 & 0.857 & 4.30 & 4.33 & 11.92 & 0.9 M \\
      & 16 &  $\checkmark$  & $\times$   & 0.983 & 0.954 & 0.871 & 3.67 & 4.30 & 10.51 & 3.6 M \\
      & 16 &  $\checkmark$  &  $\checkmark$ & \textbf{0.990} & 0.959 & 0.888 & 3.34 & 3.63 & 4.87 & 3.6 M \\ \hline
     
     DR-UNet 3D & 16 &  $\checkmark$  &  $\checkmark$ & 0.989 & 0.958 & 0.887 & 4.57 & 3.71 & 5.04 & 3.6 M \\ \hline
     \multicolumn{11}{c}{\textbf{End-to-End Training}} \\\hline
     
     DR-UNet 2D \& 3D CNN & 16 &  $\checkmark$  &  $\checkmark$ & 0.990 &\textbf{0.959} & 0.889 & 4.01 & 3.60 & 3.90 & 3.6 M \\
          
     DR-UNet 2D \& 3D spatio-temporal & 16 &  $\checkmark$  &  $\checkmark$ & 0.989 & 0.957 & 0.881 & 4.05 & 3.50 & 4.50 & 3.6 M \\
     
     DR-UNet 3D \& 3D CNN & 16 &  $\checkmark$  &  $\checkmark$ & 0.990 &0.959 & 0.883 & 3.80 & 3.35 & 4.03 & 3.6 M \\
     
    R-UNet 3D \& 3D spatio-temporal & 16 &  $\checkmark$  &  $\checkmark$ & 0.990 & 0.957 &\textbf{ 0.889} & \textbf{3.29} & \textbf{2.85 }& \textbf{3.40} & 3.6 M \\
     
     \hline
     \end{tabular}%
   \label{tab:addlabel}%
 \end{table*}%

\begin{figure*}[!t]
    \centering
    \includegraphics[width=0.92\textwidth]{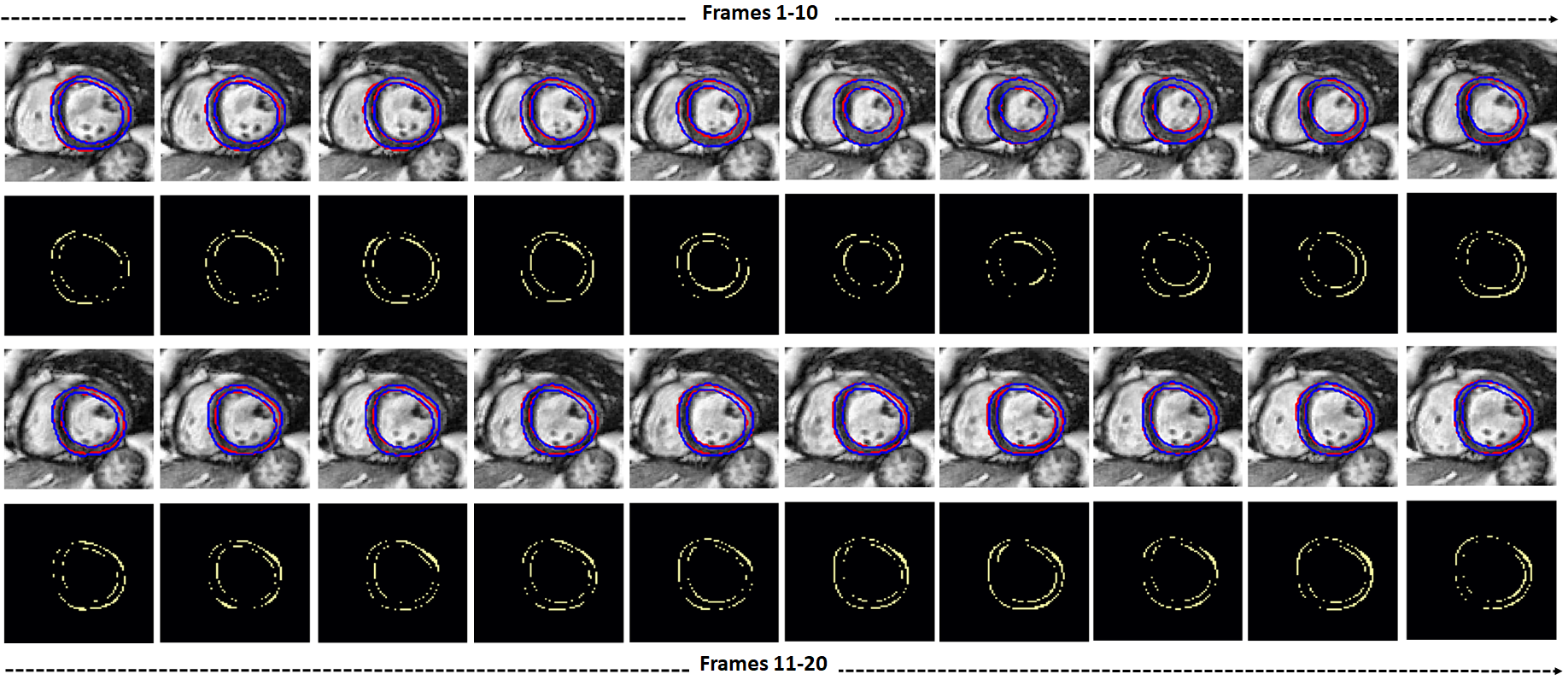}
    \caption{Qualitative segmentation results of DR-UNet 2D. The red contour shows the ground truth segmentation contour, and the blue color is overlaid as the proposed model prediction output. The row 1-2 shows the first 10 slices with their segmentation error. The row 2-4 illustrate frames 11 to 20 with segmentation error respectively. It can be seen, that segmentation error is quite low for both endocardium and epicardium contours in most of the frames.}
    \label{fig:sliceplot}
\end{figure*}
 
\begin{figure*}[ht]
    \centering
    \includegraphics[width=0.95\textwidth]{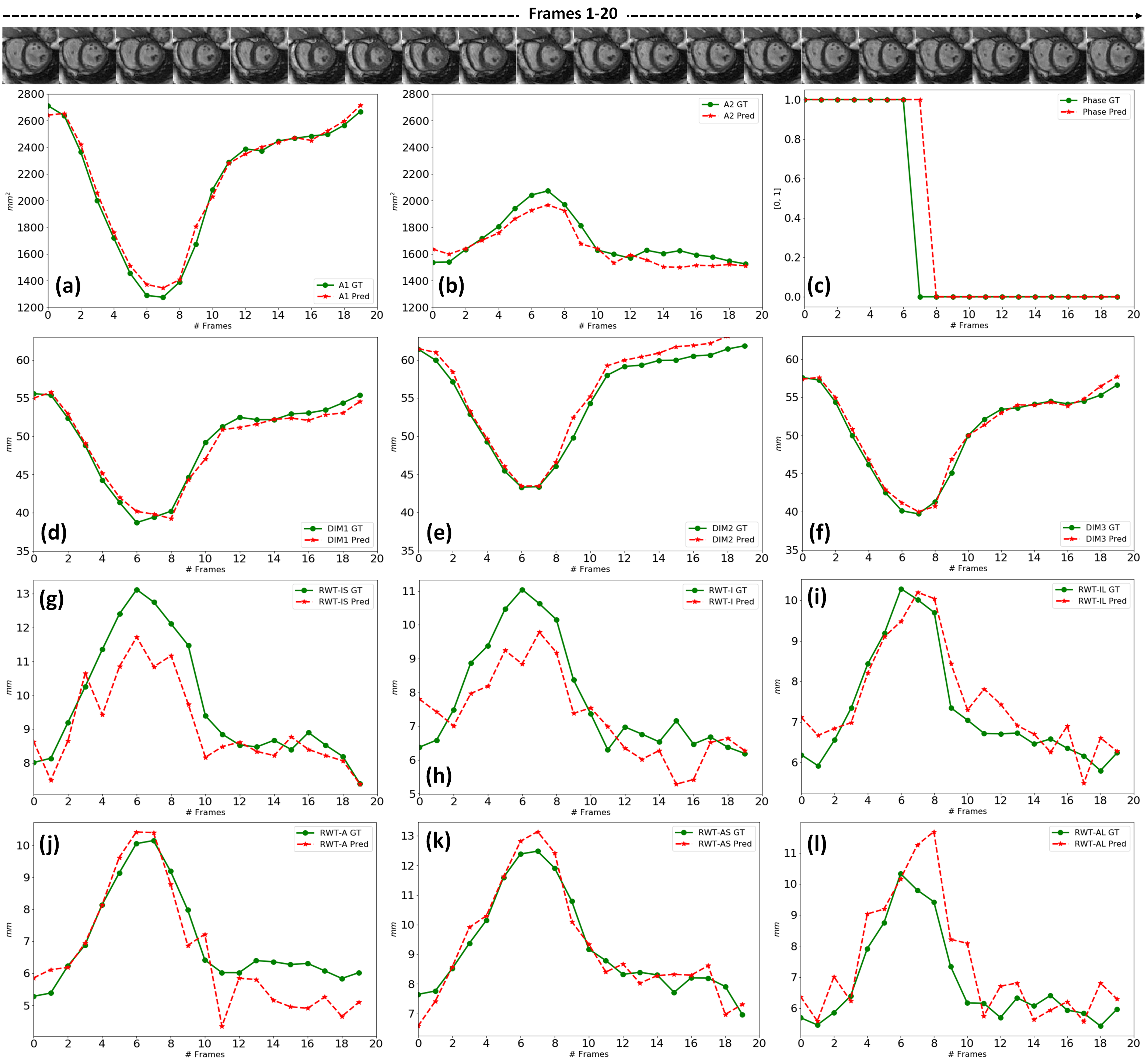}
    \caption{Example of LV indices and cardiac phase estimation by our proposed framework for a median Cine-MR case across the whole cardiac cycle. The estimated results (dashed-line in red) are very close to their ground truth values (solid line in green color) for the three types of LV indices: areas (a-b), diameters (d-f), wall-thickness (g-l) and phase cycles (c). The proposed method captures the temporal variation pattern of all indices precisely. The corresponding MR images are shown in the top row for visual comparison.}
    \label{fig:lvplot}
\end{figure*}

\begin{table*}[!t]
    \centering
    \caption{Performance of full LV quantification for existing state-of-the-art methods and our proposed methods. MAE and PCC are shown in each cell. Our method outperforms all existing methods for the three types of LV indices and cardiac phase in terms of average MAE, PCC and $ER$.}
      \resizebox{0.95\textwidth}{!}{\begin{tabular}{llcccccccc}
     \hline
     Indices & Metric & Indices-Net \cite{indnet} & FullLVNet \cite{XUE20182}  & \shortstack{FullLVNet \\(intra/inter) \cite{XUE20182}} & DMTRL \cite{XUE201854} & DLA \cite{10.1007/978-3-030-12029-0_41} & CSRNet \cite{8674807} & \shortstack{Ours \\ multi-stage} & \shortstack{Ours \\ end-to-end} \\
     \hline
     \hline
     \multicolumn{10}{c}{\textbf{Area ($mm^2$)}} \\
     \hline
     A-cav & MAE & 185$\pm$162 & 205$\pm$182 & 181$\pm$155 & 172$\pm$148 & 135 & 107$\pm$98 & 106$\pm$87 & \textbf{101$\pm$92} \\
        & PCC & 0.953 & 0.926 & 0.94 & 0.943 & \textbackslash{} & 0.982 & 0.985 & \textbf{0.986} \\
     A-myo & MAE & 223$\pm$193 & 204$\pm$195 & 199$\pm$174 & 189$\pm$159 & 177 & 162$\pm$127 & 165$\pm$132 & \textbf{158$\pm$128}\\
        & PCC & 0.853 & 0.925 & 0.935 & 0.947 & \textbackslash{} & 0.928 & 0.935 & \textbf{0.940} \\
     \textbf{Average} & MAE & 204$\pm$133 & 205$\pm$145 & 190$\pm$128 & 180$\pm$118 & 156 & 134$\pm$115 & 135$\pm$29 & \textbf{129$\pm$115} \\
        & PCC & 0.903 & 0.925 & 0.937 & 0.945 & \textbackslash{} & 0.955 & 0.960 & \textbf{0.964}\\
     \hline
     \hline
     \multicolumn{10}{c}{\textbf{Dimension ($mm$)}} \\
     \hline
     Dim1 & MAE & \textbackslash{} & 2.87$\pm$2.23 & 2.62$\pm$2.09 & 2.47$\pm$1.95 & 2.04 & \textbf{1.57$\pm$1.42} & 1.76$\pm$1.43 & 1.85$\pm$1.49\\
        & PCC & \textbackslash{} & 0.938 & 0.952 & 0.957 & \textbackslash{} & 0.974 &\textbf{ 0.975} &0.973\\
     Dim2 & MAE & \textbackslash{} & 2.96$\pm$2.35 & 2.64$\pm$2.12 & 2.59$\pm$2.07 & 2.02 & \textbf{1.48$\pm$1.36} & 1.80$\pm$1.49 & 1.76$\pm$1.44\\
        & PCC & \textbackslash{} & 0.864 & 0.881 & 0.894 & \textbackslash{} & \textbf{0.979} & 0.977 & 0.977\\
     Dim3 & MAE & \textbackslash{} & 2.92$\pm$2.48 & 2.77$\pm$2.22 & 2.48$\pm$2.34 & 2.05 & \textbf{1.56$\pm$1.33} & 1.72$\pm$1.41 & 1.82$\pm$1.43\\
        & PCC & \textbackslash{} & 0.924 & 0.935 & 0.943 & \textbackslash{} & 0.979 & \textbf{0.978} & 0.975 \\
     \textbf{Average} & MAE & \textbackslash{} & 2.92$\pm$1.89 & 2.68$\pm$1.64 & 2.51$\pm$1.58 & 2.03 & \textbf{1.54$\pm$1.37 }& 1.76$\pm$1.44 & 1.81$\pm$1.46 \\
        & PCC & \textbackslash{} & 0.901 & 0.917 & 0.925 & \textbackslash{} & 0.978 & 0.977& 0.975 \\
     \hline
     \hline
     \multicolumn{10}{c}{\textbf{RWT ($mm$)}} \\
     \hline
     IS & MAE & 1.39$\pm$1.13 & 1.42$\pm$1.21 & 1.32$\pm$1.09 & 1.26$\pm$1.04 & 1.39 & \textbf{1.06$\pm$0.87} & 1.15$\pm$0.93 & 1.16$\pm$0.921\\
        & PCC & 0.824 & 0.806 & 0.84 & 0.856 & \textbackslash{} & 0.895 & 0.908&\textbf{0.910} \\
     I  & MAE & 1.51$\pm$1.21 & 1.53$\pm$1.25 & 1.38$\pm$1.10 & 1.40$\pm$1.10 & 1.41 & 1.33$\pm$1.14 & \textbf{1.24$\pm$1.01} & 1.25$\pm$1.01 \\
        & PCC & 0.701 & 0.678 & 0.751 & 0.747 & \textbackslash{} & 0.812 & \textbf{0.856} & 0.855 \\
     IL & MAE & 1.65$\pm$1.36 & 1.74$\pm$1.43 & 1.57$\pm$1.35 & 1.59$\pm$1.29 & 1.48 & \textbf{1.33$\pm$1.09} & 1.42$\pm$1.13 & 1.47$\pm$1.17 \\
        & PCC & 0.671 & 0.618 & 0.691 & 0.693 & \textbackslash{} & 0.788 & \textbf{0.836}& 0.825 \\
     AL & MAE & 1.53$\pm$1.25 & 1.59$\pm$1.31 & 1.60$\pm$1.36 & 1.57$\pm$1.34 & 1.46 & \textbf{1.32$\pm$1.09 }& 1.37$\pm$1.08 & 1.38$\pm$1.06 \\
        & PCC & 0.698 & 0.657 & 0.651 & 0.659 & \textbackslash{} & 0.77 & \textbf{0.829}& \textbf{0.831} \\
     A  & MAE & 1.30$\pm$1.12 & 1.36$\pm$1.17 & 1.34$\pm$1.11 & 1.32$\pm$1.10 & 1.24 & \textbf{1.08$\pm$0.92} & 1.13$\pm$0.97 & 1.14$\pm$0.99 \\
        & PCC & 0.781 & 0.754 & 0.768 & 0.777 & \textbackslash{} & 0.84 & \textbf{0.875}& 0.870 \\
     AS & MAE & 1.28$\pm$1.00 & 1.43$\pm$1.24 & 1.26$\pm$1.10 & 1.25$\pm$1.01 & 1.31 & \textbf{0.97$\pm$0.80} & 1.05$\pm$0.84 & 1.03$\pm$0.83 \\
        & PCC & 0.871 & 0.821 & 0.864 & 0.877 & \textbackslash{} & 0.919 &0.928&\textbf{0.933}\\
     \textbf{Average} & MAE & 1.44$\pm$0.71 & 1.51$\pm$0.81 & 1.41$\pm$0.72 & 1.39$\pm$0.68 & 1.38 & \textbf{1.16$\pm$0.097} & 1.23$\pm$1.01 & 1.24$\pm$1.01\\
        & PCC & 0.758 & 0.723 & 0.761 & 0.768 & \textbackslash{} & 0.868 & \textbf{0.872 } & 0.871\\
     \hline
     \hline
     \multicolumn{10}{c}{\textbf{Phase ($\%$)}} \\
     \hline
     ES/DS &  ER  & \textbackslash{} & 13 & 10.4 & 8.2 & \textbf{8.1} & \textbackslash{} & 10.8 & 9.0\\
     \hline
     \hline
    \end{tabular}}%
    \label{tab:result}%
\end{table*}%

\subsection{Comparison With State-of-the-art Methods}
\noindent\textbf{LV Segmentation Performance:} All segmentation networks were evaluated with and without the connected component analysis (CCA) as postprocessing step and weighted-class loss function, respectively. Table \ref{tab:addlabel} summarizes the results for our segmentation module $\mathbf{G}$ under different settings. We observe that DR-UNet 2D and DR-UNet 3D achieved high segmentation accuracy for the LV blood pool and background classes in terms of Dice score and HD value. However, it was less successful for the Myo, primarily due to the presence of noise, low contrast tissues, and different pathologies.  
The 2D DR-UNet with eight filters without CCA and weighted class average loss achieved an average Dice of 85.0\% on the validation set for Myo. However, by including both operations and increasing the number of filters to 16, the average Dice for Myo improved to 89.0\%. HD value also reduced to 4.87 $mm$, respectively. Accurate segmentation of Myo is crucial since most LV functional indices are computed based on the endo- and epicardial contours. A similar performance gain was achieved by DR-UNet 3D. DR-UNet 3D with CCA and weighted-loss achieved an average Dice score of 88.7\% for Myo, but a substantially higher HD value of 3.71 $mm$. Fig. \ref{fig:sliceplot} depicts the segmentation output for the complete cardiac cycle of a patient and segmentation error w.r.t the ground truth. It can be seen that the predicted contours for LV and Myo are precise and have a very low error-rate at tissue boundaries. 
Table \ref{tab:addlabel} also demonstrates the intermediate segmentation results for DR-UNet 3D and DR-UNet 2D when the models were trained end-to-end along with the regression and classification tasks (rows 8-11). Interestingly, DR-UNet 3D and 3D spatio-temporal network achieved a dice score of 89.0\% and the lowest HD values of 2.85 $mm$ and 3.40 $mm$ for LV and Myo. It can also confirm the advantage afforded by jointly optimizing three tasks together under a multi-task learning scenario.
\\

\noindent\textbf{LV Quantification Performance:} In order to highlight the gain in performance for LV quantification afforded by our approach relative to the state-of-the-art, we compared our approach with six recent methods - Xue \textit{et al.} \cite{indnet} (Indices-Net) , Xue \textit{et al.} \cite{XUE20182} (FullLVNet), Xue \textit{et al.} \cite{XUE201854} (DMTRL),  Li \textit{et al.} \cite{10.1007/978-3-030-12029-0_41} (DLA), and  Wang \textit{et al.} \cite{8674807} (CSRNet) under the same experiment settings. Xue \textit{et al.} introduced Indices-Net to estimate multiple cardiac indices at the same time. The author uses two closely coupled networks: a deep convolutional auto-encoder for feature extraction from cardiac images and a multiple-output CNN for index regression. Xue \textit{et al.} extended their algorithm \cite{XUE201854} to first learn cardiac representations with a deep CNN, and subsequently, the temporal dynamics of the cardiac sequence with two parallel RNN modules. Li \textit{et al.} \cite{10.1007/978-3-030-12029-0_41} proposed a method based on deep learning that includes 11 indices of regression and cardiac phase detection. The authors use deep layer aggregation (DLA) as the backbone to perform 11 index regressions simultaneously on 2D single images and derive the cardiac phase by searching for the maximum and minimum frames from the polynomial LV cavity region. In the most recent attempt for LV quantification, Wang \textit{et al.} \cite{8674807} proposed CSRNet as an end-to-end framework that computes the LV indices based on the segmentation mask similar to our model. However, most of these only use a single 2D image or embedding of 5 images for feature extraction. 

Our proposed spatio-temporal multi-task learning approach outperformed the majority of these state-of-the-art methods for estimating LV indices, evaluated in terms of mean absolute error (MAE) and the Pearson correlation coefficient (PCC). These metrics were evaluated with respect to the ground truth values and are reported in TABLE \ref{tab:result}. We can see that our methods yield the lowest average MAE of 129$\pm$115 $mm^2$ and PCC of 0.964 for the LV blood pool and Myo areas, compared to all other methods. Moreover, it achieved an average MAE of 1.76$\pm$1.44 $mm$ and 1.24$\pm$1.01 $mm$ for the LV dimensions and RWTs, which are very close to the state-of-the-art results reported by Wang \textit{et al.} \cite{8674807} on this benchmark dataset. However, in terms of PCC values, we outperformed the model proposed by Wang \textit{et al.}, achieving values of $0.975$ and $0.871$.  On the other hand, there is a reduction of 38.8\% MAE value for the cavity area and 38.6\% for the Myo area when compared to the IndiceNet, FullLVNet, DMTRL, and DRL methods. The higher average PCC value means that there is a better linear relationship between the estimation results from our model and the ground truth, which is illustrated for a test case in Fig. \ref{fig:lvplot}. It can be seen in the figure that RWTs are more difficult to estimate in comparison to LV dimensions and area of the cavity. This is because RWTs estimation involves both the endocardium and epicardium contours (small region), which is usually difficult to segment due to available noise and low contrast within the image.  Furthermore, the improvements afforded by our approach to estimates the myocardial indices and RWTs, in terms of MAE are less prevalent as the segmentation results for the LV blood pool were more accurate than for the Myo. 


Our model is trained both in an end-to-end and multi-stage fashion. The end-to-end model unified all three modules (segmentation, regression, and classification), and based on evaluation metrics achieved overall better performance compared to the former.

\noindent\textbf{Ablation Studies:}
We conducted ablation experiments to evaluate the effectiveness of spatio-temporal convolutions in our proposed spatio-temporal multi-task framework. The results are presented in Table \ref{tab:synergil}. Our baseline network uses only 3D convolution layers to incorporate temporal information for better LV indices regression and phase classification in a multi-stage manner (row 1). It can be seen from the table, that this configuration achieved only an average MAE value of 2.06 $mm$ for LV blood pool dimensions, 154 $mm^2$ for LV and Myo areas, and 1.35$mm$ for RWTs. The error rate for phase classification is also quite high close to 11.0$\%$. Training this network with the same configuration in an end-to-end manner improved the LV areas, dimensions and RWT quantification slightly (row 2). Next, we replaced the 3D convolution blocks with proposed $2D + t$ spatio-temporal layers with a kernel size of $3 \times 1 \times 1$ and $1 \times 3 \times 3$. This model trained both in a multi-stage and end-to-end fashion (rows 5-8). We can see that the MAE and ER values improved almost for all the indices in comparison to 3D convolution configuration. The model with multi-stage training and spatio-temporal CNN achieved the lowest MAE value of 1.76 $mm$ and 1.23 $mm$  for LV dimensions and RWTs (row 7). However, the 3D spatio-temporal CNN with an end-to-end training strategy achieved the lowest MAE value for LV and Myo areas and reduced the phase classification error-rate to 9.0$\%$ (row 8). Moreover, we have also trained two models based on our DR-UNet 2D segmentation model in multi-stage and end-to-end fashion (rows 3 \& 5). Here, we can see again that DR-UNet 2D and 3D spatio-temporal model outperformed DR-UNet 2D and 3D CNN. These results can confirm the effectiveness of spatio-temporal layers, which encode jointly global temporal information and local spatial information.

 \begin{table*}[ht]
   \centering
   \caption{Effectiveness of 3D spatio-temporal CNN layers in our proposed framework in comparison to 2D and 3D convolution operations. Multi-stage and End-to-End denotes the type of training strategy. The values show the average MAE for LV blood pool and Myo areas, 3 LV dimensions and 6 RWTs.}
     \resizebox{0.80\textwidth}{!}{\begin{tabular}{|l|c|c|c|c|c|c|}
     \hline
     \textbf{Methods} & \textbf{\shortstack{Multi-stage}} & \textbf{\shortstack{End-to-End}} & \textbf{\shortstack{Area ($mm^2$)}} & \textbf{\shortstack{DIM ($mm$)}} & \textbf{\shortstack{RWT ($mm$)}} & \textbf{\shortstack{Phase (\%)}} \\
     \hline
     \hline
         DR-UNet 3D \& 3D CNN & $\checkmark$   &   $\times$   & 154$\pm$134 & 2.06$\pm$1.60 & 1.35$\pm$1.09 & 11.0   \\
     \hline
          DR-UNet 3D \& 3D CNN &  $\times$   &   $\checkmark$   & 148$\pm$122 & 2.00$\pm$1.32 & 1.32$\pm$1.02 & 11.2   \\
     \hline
          DR-UNet 2D \& 3D CNN &  $\checkmark$   &   $\times$   & 157$\pm$127 & 2.01$\pm$1.64 & 1.30$\pm$1.03 & 11.3   \\
     \hline
           DR-UNet 2D \& 3D CNN &  $\times$   &   $\checkmark$   & 146$\pm$116 & 2.16$\pm$1.73 & 1.32$\pm$1.01 & 10.6   \\
     \hline
         DR-UNet 2D \& 3D spatio-temporal &  $\checkmark$   &   $\times$   & 136$\pm$115 & 1.77$\pm$1.44 & 1.23$\pm$1.01 & 10.7  \\
     \hline
          DR-UNet 2D \& 3D spatio-temporal &  $\times$   &   $\checkmark$   & 142$\pm$113 & 2.13$\pm$1.71 & 1.25$\pm$1.05 & 10.6  \\
     \hline
     
        DR-UNet 3D \& 3D spatio-temporal & $\checkmark$    &   $\times$   & 135$\pm$29 & \textbf{1.76$\pm$1.44} & \textbf{1.23$\pm$1.01} & 10.8 \\
     \hline
         DR-UNet 3D \& 3D spatio-temporal & $\times$    &  $\checkmark$   & \textbf{129$\pm$115} & 1.81$\pm$1.46 & 1.24$\pm$1.01 & \textbf{9.0} \\
     \hline
     \end{tabular}}%
   \label{tab:synergil}%
 \end{table*}%

\subsection{Statistical Analysis}
To statistically measure the effectiveness of our proposed multi-task approach compare to manual ground-truths, we employ Bland-Altman analysis \cite{MARTINBLAND1986307}. This statistical technique determines the agreement between two quantitative measurements by constructing limits of agreement (LoA). 

\begin{figure*}[ht]
    \centering
    \subfigure[]{%
    \label{fig:first1}%
    \includegraphics[width =0.333\textwidth]{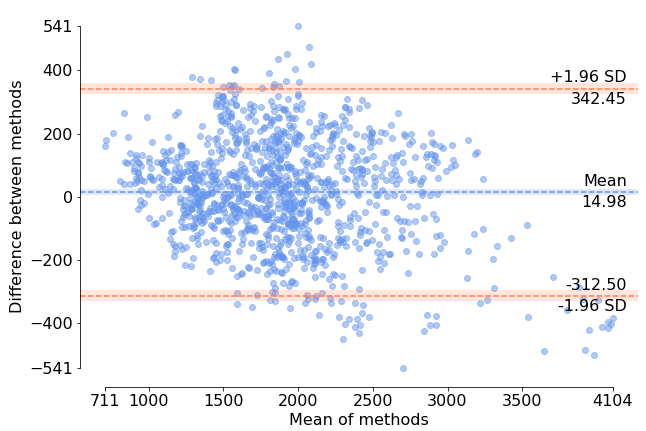}}%
    \subfigure[]{%
    \label{fig:second}%
    \includegraphics[width =0.330\textwidth]{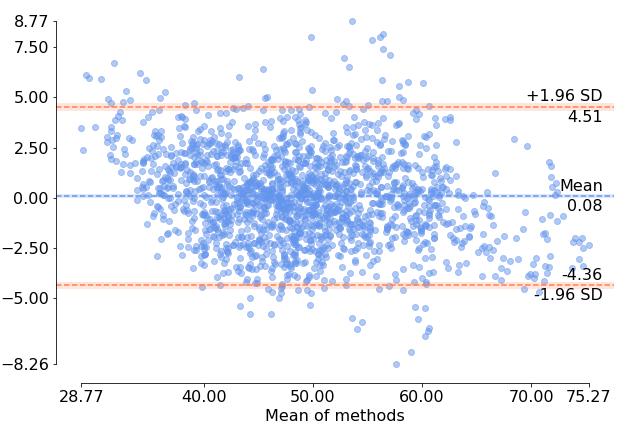}}%
    \vspace{0.00mm}
    \subfigure[]{%
    \label{fig:blandaltman}%
    \includegraphics[width =0.330\textwidth]{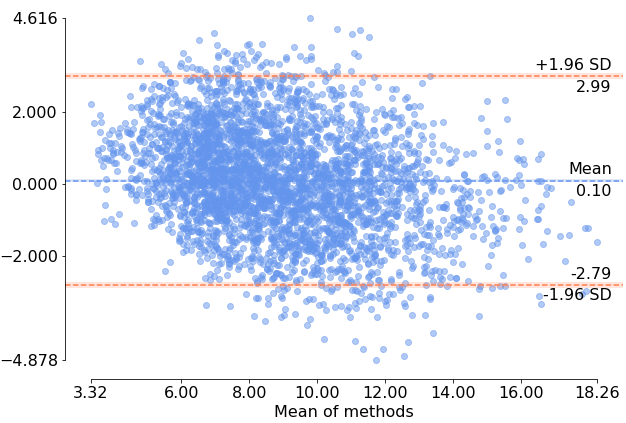}}%
    \qquad
    \caption{
    Bland-Altman analysis plots show that the LV indices for areas, dimensions, and RWTs estimated using our model is very close to the ground truth. Plot (a) illustrates the LV cavity and Myo areas (A1 and A2). The plots for LV blood pool dimensions and RWTs are shown in b-c. The Bland-Altman plots are computed using a confidence interval of 95\%. The blue line indicates the mean and the dashed red lines indicate the level of agreement.}
    \label{blandalt}
\end{figure*}

The Bland-Altman plots for differences in LV indices (Areas, DIMs, and RWTs) obtained using manual and proposed methods are shown in Fig. \ref{blandalt}. The areas per patient are expressed in $mm^{2}$ and for dimensions and RWTs in $mm$ respectively. It can be observed that in terms of LV area estimation (Fig. \ref{fig:first1}) the agreement between our proposed method and manually generated ground truth is high with a bias (mean signed difference) of 14.98 $mm^{2}$ and limits of agreement of $\pm$342.45 $mm^{2}$. This is also the same for the dimensions and RWTs with LoR of $\pm$4.51 $mm^{2}$ and $\pm$2.99 $mm^{2}$, respectively. These results suggest that the proposed method has a small bias to overestimate RWTs and that the variation between automated and manual estimates of the LV area is only slightly greater than the expert manual annotation. There are some outlier cases (refer to plots a-c in Fig. \ref{blandalt}), regarded as hard-examples to measure due to the presence of low contrast and noise in the scans or not a precise segmentation of LV and Myo cavity areas. Overall, our methods produced accurate indices quantification resulting in a significantly lower mean difference in most of the cases.

\section{Discussion}
We proposed a novel multi-task end-to-end method for full LV quantification in cardiac cine-MR images in this study. This approach leverages spatial and temporal information contained within the cine-MR sequences using 3D spatio-temporal convolutions, to quantify the LV, unlike most existing methods that utilize just 2D spatial convolutions. Additionally, our model exploits the information contained in the estimated segmentation masks rather than the raw images, to combine both spatial and temporal features during model training, and inference. Thus, features are learned from the full cine-MR sequence to estimate the various LV indices of interest. This framework can be considered as an efficient tool for cardiac LV functional analysis that tackles three different related tasks simultaneously, namely - LV blood pool and Myo segmentation, regression of 11 LV morphological indices, and classification of cardiac phase.

A multi-task learning framework enables specialized modules to tackle different tasks simultaneously while benefiting from one another. By sharing the same feature extraction backbone, this framework allows information synergy between various tasks and presents a mutual influence process that can further obtain performance gains from different tasks. In our network, the regression and classification tasks are optimized in an end-to-end paradigm, together with LV segmentation. The presented results indicate that methodically consolidating multiple but interrelated tasks with mutual information sharing and considering the task relationship using a suitable weighting strategy, yields better performance. Moreover, our method is similar to the commensal correlation network proposed by Luo \textit{et al.} \cite{LUO2020101591}, where feature extraction performed in parallel to the segmentation for LV quantification. However, this method computes the LV indices using a single 2D MR image and temporal information of the cardiac cycle discarded. In contrast, our proposed method takes advantage of the full cardiac cycle and the quantification indices computed in a cascade manner.  

Extensive experiments on the LVQuan 2018 benchmark dataset have highlighted the effectiveness of our approach. By integrating myocardial segmentation with multi-task classification and regression learning framework. The method combines the advantages of two-step methods based on segmentation with end-to-end learning approaches. The segmentation module of the framework can remove task-independent structures so that the following regression and classification network can extract discriminating features from the segmented masks only. The myocardial contours only guide the regression task, but do not fully determine the accuracy of the quantification results like the two-step procedures based on segmentation. The results for DR-UNet with post-processing and weighted-loss outperformed the other segmentation methods. Additionally, our segmentation network has fewer than 3.6 million trainable parameters and takes less than 20 minutes to train with 300 epochs, which is considered modest in size/complexity, compared with other relevant SOTA approaches \cite{XUE201854}. In the validation phase, LV metrics are estimated in each fold for 28 subjects with 580 images in just $\sim$2.2 seconds on a machine with 4GB GPU memory. This demonstrates the real-time characteristic of the pipeline, which could be integrated into MR acquisition systems to triage patients into high and low-risk categories of CVDs for improved efficiency and clinical decision making.

The proposed method achieved comparable results to CSRNet and even for LV and Myo areas, our model achieved better results. On the other hand, for 3 LV diameters and 6 RWT, our model could achieve very close results. However, It should be noted that CSRNet doesn't consider cardiac phase detection tasks (systolic or diastolic) and only quantify the indices based on raw segmentation. On the other hand, our method simultaneously performs three tasks including segmentation, indices regression and phase classification. In this end-to-end training, every learning step is directed at the final goal, encoded by the overall objective function. There is no need for training modules on an auxiliary objective. Our end-to-end multi-task learning framework is nicely consistent with the general approach of machine learning to take the human expert out of the loop and to solve problems in a purely data-driven manner.

Although the segmentation accuracy of our approach was high for the LV blood pool, there is still room for improvement, especially for the myocardium. Moreover, in the end-to-end training strategy, combining the losses of different tasks is a critical issue, because different tasks converge at different rates. To balance task importance during optimization, we empirically set the hyperparameter $\lambda$ for each task in the loss function after a greedy search. We believe incorporating a more methodical multi-task loss weighting strategy could improve the performance of our pipeline even further. Furthermore, the LVQuan benchmark dataset is preprocessed \textit{a priori}, and LV ROIs are extracted from Cine-MR sequences, which is not common in a real clinical scenario. We aim to extend our method to include LV detection within the pipeline and design a fully automated computer-aided diagnosis system that eliminates the need for any pre-processing step. In this study, all quantification indices are 2D, while the most important cardiac functional/morphological indices are 3D metrics like ejection fraction and LV volumes. As future work, we also aim to extend our method to include these metrics as well.

\section{Conclusion}
In this study, we proposed a robust, efficient, and light-weight network architecture for fully automatic LV quantification in Cine-MR images. The proposed network first segments the LV and Myo blood pool. Subsequently, the segmented structures fed to a spatio-temporal multi-task regression and classification component to estimate cardiac LV indices. It includes the LV cavity and Myo areas, LV dimensions, myocardial RTWs, and the cardiac cycle phase (diastolic or systolic). Although the LV anatomical shape and appearance are highly variable across different subjects and the training subjects acquired in various hospitals, the proposed method successfully learned robust representations from the MR sequences and estimated the LV indices of interest with high accuracy. We evaluated our method on 145 subjects, and the experimental results highlighted the advantage afforded by our approach in comparison to the state-of-the-art methods. The proposed method can be a promising contribution having clinical importance both during diagnosis, where the cardiac MR volume needs to be analyzed and during treatment planning when the quantification of the anatomical structure of LV needs to be accurate and fast.

%


\ifCLASSOPTIONcaptionsoff
  \newpage
\fi



%


\bibliographystyle{IEEEtran}
\bibliography{bare_jrnl_short.bib}

\begin{thebibliography}{10}
\providecommand{\url}[1]{#1}
\csname url@samestyle\endcsname
\providecommand{\newblock}{\relax}
\providecommand{\bibinfo}[2]{#2}
\providecommand{\BIBentrySTDinterwordspacing}{\spaceskip=0pt\relax}
\providecommand{\BIBentryALTinterwordstretchfactor}{4}
\providecommand{\BIBentryALTinterwordspacing}{\spaceskip=\fontdimen2\font plus
\BIBentryALTinterwordstretchfactor\fontdimen3\font minus
  \fontdimen4\font\relax}
\providecommand{\BIBforeignlanguage}[2]{{%
\expandafter\ifx\csname l@#1\endcsname\relax
\typeout{** WARNING: IEEEtran.bst: No hyphenation pattern has been}%
\typeout{** loaded for the language `#1'. Using the pattern for}%
\typeout{** the default language instead.}%
\else
\language=\csname l@#1\endcsname
\fi
#2}}
\providecommand{\BIBdecl}{\relax}
\BIBdecl

\bibitem{10.1093/eurheartj/ehx628}
E.~S.~D. Group, A.~Timmis, E.~Wilkins, L.~Wright, N.~Townsend \emph{et~al.},
  ``{European Society of Cardiology: Cardiovascular Disease Statistics 2017},''
  \emph{European Heart Journal}, vol.~39, no.~7, pp. 508--579, 11 2017.

\bibitem{doi:10.1161/CIR.0000000000000485}
E.~J. Benjamin, M.~J. Blaha, S.~E. Chiuve, M.~Cushman, S.~R. Das \emph{et~al.},
  ``Heart disease and stroke statistics\&\#x2014;2017 update: A report from the
  american heart association,'' \emph{Circulation}, vol. 135, no.~10, pp.
  e146--e603, 2017.

\bibitem{doi:10.1177/2048004016687211}
J.~Stewart, G.~Manmathan, and P.~Wilkinson, ``Primary prevention of
  cardiovascular disease: A review of contemporary guidance and literature,''
  \emph{JRSM Cardiovascular Disease}, vol.~6, p. 2048004016687211, 2017.

\bibitem{doi:10.1002/mrm.26631}
M.~R. Avendi, A.~Kheradvar, and H.~Jafarkhani, ``Automatic segmentation of the
  right ventricle from cardiac mri using a learning-based approach,''
  \emph{Magnetic Resonance in Medicine}, vol.~78, no.~6, pp. 2439--2448, 2017.

\bibitem{KIM20091}
H.~W. Kim, A.~Farzaneh-Far, and R.~J. Kim, ``Cardiovascular magnetic resonance
  in patients with myocardial infarction: Current and emerging applications,''
  \emph{Journal of the American College of Cardiology}, vol.~55, no.~1, pp. 1
  -- 16, 2009.

\bibitem{doi:10.1161/CIRCIMAGING.117.007165}
M.~Cantinotti and M.~Koestenberger, ``Quantification of left ventricular size
  and function by 2-dimensional echocardiography: So basic and so difficult,''
  \emph{Circulation: Cardiovascular Imaging}, vol.~10, no.~11, p. e007165,
  2017.

\bibitem{Bai2018}
W.~Bai, M.~Sinclair, G.~Tarroni, O.~Oktay, M.~Rajchl \emph{et~al.}, ``Automated
  cardiovascular magnetic resonance image analysis with fully convolutional
  networks,'' \emph{Journal of Cardiovascular Magnetic Resonance}, vol.~20,
  no.~1, p.~65, Sep 2018.

\bibitem{kurzendorfer}
T.~Kurzendorfer, C.~Forman, M.~Schmidt, C.~Tillmanns, A.~Maier \emph{et~al.},
  ``{Fully automatic segmentation of left ventricular anatomy in 3-DLGE-MRI},''
  \emph{Computerized Medical Imaging and Graphics}, vol.~39, no.~59, pp.
  13--27, 2017.

\bibitem{10.1007/978-3-642-33418-4_66}
M.~Afshin, I.~B. Ayed, A.~Islam, A.~Goela, T.~M. Peters \emph{et~al.}, ``Global
  assessment of cardiac function using image statistics in mri,'' in
  \emph{Medical Image Computing and Computer-Assisted Intervention -- MICCAI
  2012}.\hskip 1em plus 0.5em minus 0.4em\relax Berlin, Heidelberg: Springer
  Berlin Heidelberg, 2012, pp. 535--543.

\bibitem{10.1093/ehjci/jev014}
R.~M. Lang, L.~P. Badano, V.~Mor-Avi, J.~Afilalo, A.~Armstrong \emph{et~al.},
  ``{Recommendations for Cardiac Chamber Quantification by Echocardiography in
  Adults: An Update from the American Society of Echocardiography and the
  European Association of Cardiovascular Imaging},'' \emph{European Heart
  Journal - Cardiovascular Imaging}, vol.~16, no.~3, pp. 233--271, 02 2015.

\bibitem{6650070}
M.~{Afshin}, I.~B. {Ayed}, K.~{Punithakumar}, M.~{Law}, A.~{Islam}
  \emph{et~al.}, ``Regional assessment of cardiac left ventricular myocardial
  function via mri statistical features,'' \emph{IEEE Transactions on Medical
  Imaging}, vol.~33, no.~2, pp. 481--494, Feb 2014.

\bibitem{Tao2018}
Q.~Tao, W.~Yan, Y.~Wang, E.~H.~M. Paiman, D.~P. Shamonin \emph{et~al.}, ``Deep
  learning–based method for fully automatic quantification of left ventricle
  function from cine mr images: A multivendor, multicenter study,''
  \emph{Radiology}, vol. 290, no.~1, pp. 81--88, 2019, pMID: 30299231.

\bibitem{8674807}
W.~{Wang}, Y.~{Wang}, Y.~{Wu}, T.~{Lin}, S.~{Li} \emph{et~al.},
  ``Quantification of full left ventricular metrics via deep regression
  learning with contour-guidance,'' \emph{IEEE Access}, vol.~7, pp.
  47\,918--47\,928, 2019.

\bibitem{Suinesiaputra2015}
A.~Suinesiaputra, D.~A. Bluemke, B.~R. Cowan, M.~G. Friedrich, C.~M. Kramer
  \emph{et~al.}, ``Quantification of lv function and mass by cardiovascular
  magnetic resonance: multi-center variability and consensus contours,''
  \emph{Journal of Cardiovascular Magnetic Resonance}, vol.~17, no.~1, p.~63,
  2015.

\bibitem{RUIJSINK2019}
B.~Ruijsink, E.~Puyol-Antón, I.~Oksuz, M.~Sinclair, W.~Bai \emph{et~al.},
  ``Fully automated, quality-controlled cardiac analysis from cmr: Validation
  and large-scale application to characterize cardiac function,'' \emph{JACC:
  Cardiovascular Imaging}, 2019.

\bibitem{ATTAR201926}
R.~Attar, M.~Pereañez, A.~Gooya, X.~Albà, L.~Zhang \emph{et~al.},
  ``Quantitative cmr population imaging on 20,000 subjects of the uk biobank
  imaging study: Lv/rv quantification pipeline and its evaluation,''
  \emph{Medical Image Analysis}, vol.~56, pp. 26 -- 42, 2019.

\bibitem{BENAYED201287}
I.~B. Ayed, H.~mei Chen, K.~Punithakumar, I.~Ross, and S.~Li, ``Max-flow
  segmentation of the left ventricle by recovering subject-specific
  distributions via a bound of the bhattacharyya measure,'' \emph{Medical Image
  Analysis}, vol.~16, no.~1, pp. 87 -- 100, 2012.

\bibitem{ZHEN2016120}
X.~Zhen, Z.~Wang, A.~Islam, M.~Bhaduri, I.~Chan \emph{et~al.}, ``Multi-scale
  deep networks and regression forests for direct bi-ventricular volume
  estimation,'' \emph{Medical Image Analysis}, vol.~30, pp. 120 -- 129, 2016.

\bibitem{indnet}
W.~{Xue}, A.~{Islam}, M.~{Bhaduri}, and S.~{Li}, ``Direct multitype cardiac
  indices estimation via joint representation and regression learning,''
  \emph{IEEE Transactions on Medical Imaging}, vol.~36, no.~10, pp. 2057--2067,
  Oct 2017.

\bibitem{10.1007/978-3-319-59050-9_40}
W.~Xue, I.~B. Nachum, S.~Pandey, J.~Warrington, S.~Leung \emph{et~al.},
  ``Direct estimation of regional wall thicknesses via residual recurrent
  neural network,'' in \emph{Information Processing in Medical Imaging}, 2017,
  pp. 505--516.

\bibitem{661186}
{Chenyang Xu} and J.~L. {Prince}, ``Snakes, shapes, and gradient vector flow,''
  \emph{IEEE Transactions on Image Processing}, vol.~7, no.~3, pp. 359--369,
  March 1998.

\bibitem{spatio}
A.~Debus and E.~Ferrante, ``Left ventricle quantification through
  spatio-temporal cnns,'' in \emph{Statistical Atlases and Computational Models
  of the Heart. Atrial Segmentation and LV Quantification Challenges}, 2019,
  pp. 466--475.

\bibitem{10.1007/978-3-030-12029-0_41}
J.~Li and Z.~Hu, ``Left ventricle full quantification using deep layer
  aggregation based multitask relationship learning,'' in \emph{Statistical
  Atlases and Computational Models of the Heart. Atrial Segmentation and LV
  Quantification Challenges}, 2019, pp. 381--388.

\bibitem{XUE201854}
W.~Xue, G.~Brahm, S.~Pandey, S.~Leung, and S.~Li, ``Full left ventricle
  quantification via deep multitask relationships learning,'' \emph{Medical
  Image Analysis}, vol.~43, pp. 54 -- 65, 2018.

\bibitem{6708423}
Z.~{Wang}, M.~B. {Salah}, B.~{Gu}, A.~{Islam}, A.~{Goela} \emph{et~al.},
  ``Direct estimation of cardiac biventricular volumes with an adapted bayesian
  formulation,'' \emph{IEEE Transactions on Biomedical Engineering}, vol.~61,
  no.~4, pp. 1251--1260, April 2014.

\bibitem{XUE20182}
W.~Xue, A.~Lum, A.~Mercado, M.~Landis, J.~Warrington \emph{et~al.}, ``Full
  quantification of left ventricle via deep multitask learning network
  respecting intra- and inter-task relatedness,'' in \emph{Medical Image
  Computing and Computer Assisted Intervention − MICCAI 2017}, 2017, pp.
  276--284.

\bibitem{Meng2019}
Q.~Meng, N.~Pawlowski, D.~Rueckert, and B.~Kainz, ``Representation
  disentanglement for multi-task learning with application to fetal
  ultrasound,'' in \emph{Smart Ultrasound Imaging and Perinatal, Preterm and
  Paediatric Image Analysis}.\hskip 1em plus 0.5em minus 0.4em\relax Cham:
  Springer International Publishing, 2019, pp. 47--55.

\bibitem{CHARTSIAS2019101535}
A.~Chartsias, T.~Joyce, G.~Papanastasiou, S.~Semple, M.~Williams \emph{et~al.},
  ``Disentangled representation learning in cardiac image analysis,''
  \emph{Medical Image Analysis}, vol.~58, p. 101535, 2019.

\bibitem{ZHANG201910}
J.~Zhang, Y.~Xie, Q.~Wu, and Y.~Xia, ``Medical image classification using
  synergic deep learning,'' \emph{Medical Image Analysis}, vol.~54, pp. 10 --
  19, 2019.

\bibitem{NIPS2004_2638}
M.~Osadchy, M.~L. Miller, and Y.~L. Cun, ``Synergistic face detection and pose
  estimation with energy-based models,'' in \emph{Advances in Neural
  Information Processing Systems 17}, L.~K. Saul, Y.~Weiss, and L.~Bottou,
  Eds.\hskip 1em plus 0.5em minus 0.4em\relax MIT Press, 2005, pp. 1017--1024.

\bibitem{Maier2019}
A.~K. Maier, C.~Syben, B.~Stimpel, T.~W{\"u}rfl, M.~Hoffmann \emph{et~al.},
  ``Learning with known operators reduces maximum error bounds,'' \emph{Nature
  Machine Intelligence}, vol.~1, no.~8, pp. 373--380, 2019.

\bibitem{doi:10.1161/hc0402.102975}
M.~D. Cerqueira, N.~J. Weissman, V.~Dilsizian, A.~K. Jacobs, S.~Kaul
  \emph{et~al.}, ``Standardized myocardial segmentation and nomenclature for
  tomographic imaging of the heart,'' \emph{Circulation}, vol. 105, no.~4, pp.
  539--542, 2002.

\bibitem{10.1007/978-3-030-12029-0_35}
S.~Vesal, N.~Ravikumar, and A.~Maier, ``Dilated convolutions in neural networks
  for left atrial segmentation in 3d gadolinium enhanced-mri,'' in
  \emph{Statistical Atlases and Computational Models of the Heart. Atrial
  Segmentation and LV Quantification Challenges}, 2019, pp. 319--328.

\bibitem{resDeep}
K.~{He}, X.~{Zhang}, S.~{Ren}, and J.~{Sun}, ``Deep residual learning for image
  recognition,'' in \emph{2016 IEEE Conference on Computer Vision and Pattern
  Recognition (CVPR)}, June 2016, pp. 770--778.

\bibitem{wang2018smoothed}
Z.~Wang and S.~Ji, ``Smoothed dilated convolutions for improved dense
  prediction,'' in \emph{Proceedings of the 24th ACM SIGKDD International
  Conference on Knowledge Discovery \& Data Mining}.\hskip 1em plus 0.5em minus
  0.4em\relax ACM, 2018, pp. 2486--2495.

\bibitem{8578773}
D.~{Tran}, H.~{Wang}, L.~{Torresani}, J.~{Ray}, Y.~{LeCun} \emph{et~al.}, ``A
  closer look at spatiotemporal convolutions for action recognition,'' in
  \emph{2018 IEEE/CVF Conference on Computer Vision and Pattern Recognition},
  June 2018, pp. 6450--6459.

\bibitem{10.1007/978-3-642-15567-3_11}
G.~W. Taylor, R.~Fergus, Y.~LeCun, and C.~Bregler, ``Convolutional learning of
  spatio-temporal features,'' in \emph{Computer Vision -- ECCV 2010}.\hskip 1em
  plus 0.5em minus 0.4em\relax Berlin, Heidelberg: Springer Berlin Heidelberg,
  2010, pp. 140--153.

\bibitem{univis91902143}
S.~S. Yoon, E.~Hoppe, M.~Schmidt, C.~Forman, P.~Sharma \emph{et~al.},
  ``{Automatic Cardiac Resting Phase Detection for Static Cardiac Imaging Using
  Deep Neural Networks},'' in \emph{{Proceedings of the Joint Annual Meeting
  ISMRM-ESMRMB (27th Annual Meeting \& Exhibition)}}, I.~S. for Magnetic
  Resonance~in Medicine, Ed., 2019.

\bibitem{He}
K.~{He}, X.~{Zhang}, S.~{Ren}, and J.~{Sun}, ``Delving deep into rectifiers:
  Surpassing human-level performance on imagenet classification,'' in
  \emph{2015 IEEE International Conference on Computer Vision (ICCV)}, Dec
  2015, pp. 1026--1034.

\bibitem{Zuiderveld}
K.~Zuiderveld, ``Contrast limited adaptive histogram equalization,''
  \emph{Graphics Gems IV}, pp. 474--485, 1994.

\bibitem{7785132}
F.~{Milletari}, N.~{Navab}, and S.~{Ahmadi}, ``V-net: Fully convolutional
  neural networks for volumetric medical image segmentation,'' in \emph{2016
  Fourth International Conference on 3D Vision (3DV)}, Oct 2016, pp. 565--571.

\bibitem{Tensoflow}
M.~Abadi, P.~Barham, J.~Chen, Z.~Chen, A.~Davis \emph{et~al.}, ``Tensorflow: A
  system for large-scale machine learning,'' in \emph{12th USENIX Symposium on
  Operating Systems Design and Implementation (OSDI 16)}, 2016, pp. 265--283.

\bibitem{Adam}
D.~P. Kingma and J.~Ba, ``Adam: A method for stochastic optimization,'' 2014,
  cite arxiv:1412.6980Comment: Published as a conference paper at the 3rd
  International Conference for Learning Representations, San Diego, 2015.

\bibitem{MARTINBLAND1986307}
J.~M. Bland and D.~Altman, ``Statistical methods for assessing agreement
  between two methods of clinical measurement,'' \emph{The Lancet}, vol. 327,
  no. 8476, pp. 307 -- 310, 1986.

\bibitem{LUO2020101591}
G.~Luo, S.~Dong, W.~Wang, K.~Wang, S.~Cao \emph{et~al.}, ``Commensal
  correlation network between segmentation and direct area estimation for
  bi-ventricle quantification,'' \emph{Medical Image Analysis}, vol.~59, p.
  101591, 2020.

\end{thebibliography}

\end{document}